\documentclass[
 reprint,
superscriptaddress,
 amsmath,amssymb,
prb,
citeautoscript]{revtex4-2}

\usepackage{graphicx}
\usepackage{dcolumn}
\usepackage{bm}

\usepackage[bookmarks=true,colorlinks=true,urlcolor=blue,linkcolor=blue,citecolor=blue,breaklinks=true]{hyperref}
\PassOptionsToPackage{hyphens}{url}
\usepackage[utf8]{inputenc}
\usepackage{amsmath, amssymb}
\usepackage[table,dvipsnames]{xcolor}       

\usepackage{lipsum}
\usepackage{soul}

\newcommand{\DIPC}[0]{
Donostia International Physics Center (DIPC),
Paseo Manuel de Lardizabal 4, 20018 Donostia-San Sebasti\'an, Spain}
\newcommand{\CFM}[0]{
Centro de F\'{\i}sica de Materiales CFM/MPC (CSIC-UPV/EHU), Paseo Manuel de Lardizabal 5, 20018 Donostia-San Sebasti\'an, Spain}
\newcommand{\PolymerEHU}[0]{Departamento de Pol\'{i}meros y Materiales Avanzados: F\'{i}sica, Qu\'{i}mica y Tecnolog\'{i}a, Facultad de Qu\'{i}micas (UPV/EHU), Apartado 1072, 20080 Donostia-San Sebasti\'{a}n, Spain}

\begin{document}

\title{Nonadiabatic vibrational dynamics of CO on Pt(111) studied with first-principles many-body perturbation theory}

%
\author{Raúl Bombín}
\email{raul.bombin@ehu.eus}
\affiliation{\CFM}

\author{Iñaki Fernández Tena}
\affiliation{\CFM}

\author{J. I\~{n}aki Juaristi}
\email{josebainaki.juaristi@ehu.eus}
\affiliation{\PolymerEHU}
\affiliation{\CFM}
\affiliation{\DIPC}
\author{Maite Alducin}
\email{maite.alducin@ehu.eus}
\affiliation{\CFM}
\affiliation{\DIPC}

\date{\today}

\begin{abstract}
We investigate the nonequilibrium vibrational dynamics of CO adsorbed on Pt(111), a prototypical adsorbate–surface system for ultrafast vibrational spectroscopy, using a first-principles framework that combines many-body perturbation theory with density functional theory. Within this methodology we explicitly account for electron-hole pair excitations and electron-mediated phonon-phonon coupling. For the latter mechanism, we evaluate the vertex-correction screening, what constitutes a methodological novelty. Our calculations qualitatively reproduce the transient evolution of the internal stretch  vibrational mode observed in femtosecond pump-probe experiments, allowing us to identify in the experimental measurements a small transient blueshift after excitation with a femtosecond laser pulse. 
As predicted in a previous work, the reduction in the number of electronic states available for electron-phonon coupling when the electronic temperature raises at a group 10 transition metal surface, effectively screens the electron-phonon interaction giving rise to an anomalous transient blue-shift.
\end{abstract}

\maketitle

\section{Introduction}

The study of ultrafast elementary motions of adsorbates on metal surfaces is directed to obtain a precise and fundamental knowledge of the timescales of surface reactions. In that respect, femtosecond laser pulses constitute an efficient tool to initiate reactions and to gain information in the ultrashort time domain. More precisely, in pump-probe infrared vibrational spectroscopy experiments, the dynamics of dipolar molecular adsorbates, which is induced by the initial pump pulse, is probed by tracking the linewidth change and frequency shift of the infrared-active internal stretch (IS) mode~\cite{arnolds2010}. In most of the studied systems, after the pump pulse heats the surface, the IS mode frequency experiences an initial rapid redshift followed by a slower (but still rapid) blueshift~\cite{Germer94,Bonn2000,Lane2006,Lane2007,Omiya2014,Inoue2016}. For many years, the microscopic mechanisms ruling these spectral changes were not fully understood. In some works the observed behavior was attributed to the anharmonic coupling with other low-energy modes~\cite{Bonn2000},
and in others to the energy transfer from the pump pulse excited electrons to the adsorbates via nonadiabatic coupling (NC)~\cite{Lane2006,Inoue2016}.

A similar situation occurred regarding the linewidth and frequency shifts of the IS mode of molecules adsorbed on metals measured in infrared reflection absorption spectroscopy (IRAS) experiments under thermal conditions~\cite{Germer94,cook97,Cook1997,Beckerle1991,Schweizer1989,Carrasco10,Hoffmann86,Symonds04,Lane2007,Person1985,Roeterdink2004,Schweizer1989}. On the one hand, the linewidth values were considered fingerprints of the nonadiabatic electron-vibrational mode coupling~\cite{Ryberg85,Morin1992}. On the other hand, their temperature dependence was ascribed to the coupling with thermally excited low energy vibrational modes~\cite{Person1985}. Nevertheless, the lack of a quantitative theory based on first principles did not allow to provide a definite answer regarding the role played by the different mechanisms. 
These problems were finally solved by the introduction of a new theory that, combining many-body perturbation theory (MBPT) with density functional theory (DFT) and going to second-order in the electron-phonon coupling, achieved excellent agreement with experiments for the value of the linewidth of the IS mode of CO adsorbed on Cu(100) and its temperature dependence~\cite{Novko2018}. Briefly, apart from the contribution of the direct interband transitions obtained from the calculation of the phonon self-energy due to electron-phonon coupling up to first-order, a new term was included by calculating the second-order in the electron-phonon coupling intraband phonon self-energy. This additional term accounted for the coupling of the high-energy IS mode to low energy phonon modes, the energy mismatch being compensated by the excitation of electron-hole pairs. The new mechanism was, consistently, named electron-mediated phonon-phonon coupling (EMPPC) and was shown to be the main responsible of the temperature dependence of the IS linewidth. 

The subsequent extension of the theory to nonequilibrium conditions reproduced and successfully explained the time evolution of the linewidth and frequency changes of the CO IS mode on Cu(100) in femtosecond pump-probe infrared spectroscopy~\cite{Novko2019}. In short, during the first picosecond the frequency shift --specifically, its initial redshift followed by a blueshift-- was governed by the first-order NC term, whereas the EMPPC term was the main responsible of the time evolution of the linewidth. 
More recently, some of us applied the above formalism to the study of both the temperature dependence of the linewidth and frequency shift of the IS mode of CO on Pd(111) under thermal conditions, and their time evolution under laser-induced nonequilibrium conditions~\cite{Bombin2023letter,Bombin2023}. Interestingly, upon laser excitation the model predicted an unusual blueshift of the CO IS mode within the subpicosecond time scale~\cite{Bombin2023letter}, instead of the initial redshift observed in CO/Cu(100)~\cite{Inoue2016,Novko2019} and other systems~\cite{Bonn2000,Lane2006,Lane2007}. Moreover, it was shown that this behavior was not related to intermode coupling, but to the direct coupling of the IS mode to the laser excited hot electrons. In metals, the electron-phonon coupling strength is expected to be enhanced upon laser excitation, which causes the usual softening of the frequency. However, it was demonstrated that due to the abrupt decrease of the density of states around the Fermi level in Pd, upon laser excitation, the electron-phonon coupling is effectively screened leading to the anomalous transient blueshift. The absence of pump-probe vibrational spectroscopy experiments 
in this system did not allow to test this theoretical prediction.

In the present work, we study the CO on Pt(111) system. Since the electronic density of states of Pt is similar to that of Pd, we expect a similar behavior of the CO IS mode linewidth changes and frequency shifts both at thermal and at nonequilibrium conditions, including the above discussed unusual frequency blueshift. In contrast to CO/Pd(111), for CO/Pt(111) experimental information exists~\cite{Beckerle1991,Fournier2004,Watanabe2010,Inoue2012,Roeterdink2004,Schweizer1989,Crosslye1977,Malik1989,Hoge1988}, that will allow us to validate the predictions of the theory. 

With regard to the methodology, we introduce two improvements compared to previous studies. 
First, it should be noted that, in practice, a metal surface is necessarily described in DFT calculations with a finite slab, which in this study consists of six atomic layers. In previous works, all the optical modes of the (model) system were taken into account when evaluating the EMPPC term~\cite{Bombin2023letter,Bombin2023}. This meant that the contribution of the characteristic modes of the slab, which should not be present in the surface description, was included in the calculation. The latter can give rise to an overestimation of the EMPPC term. In the present work this issue is corrected by neglecting the contribution of slab specific modes. The second improvement involves the vertex correction in the EMPPC term. In previous works, this vertex correction was neglected in order to make the numerical simulations more feasible~\cite{Novko2018,Novko2019,Bombin2023letter,Bombin2023}. The physical justification for this relied in Migdal's theorem, which suggests that vertex corrections should not alter the result significantly because the corresponding term scales with the square root of the ratio of the electronic to the ionic mass~\cite{Migdal1958}. Here we do evaluate the exact expression for the EMPPC term, including the vertex correction, and check whether the above assumption is justified.

The organization of the paper is as follows. The theoretical method is summarized in Sec.~\ref{sec:methods}. Section~\ref{subsec:system} is devoted to the description of the CO/Pt(111) system. The results for the CO IS mode linewidth change and frequency shift under thermal conditions and under ultrashort laser pulse irradiation are presented and discussed in Secs.~\ref{sec:thermal} and~\ref{sec:ultrafast}, respectively. All our results are compared to experiments when available. Finally, the main conclusions are summarized in Sec.~\ref{sec:conclusions}.

\section{Methods}
\label{sec:methods}

The Pt(111) surface is modeled as a six-layer slab with a $(\sqrt{3}\times \sqrt{3}){\rm R}30^\circ$ surface unit cell including one CO molecule adsorbed at top position to achieve the desired coverage of 0.33~ML. 
The electronic structure is evaluated at DFT level using the \textsc{quantum espresso} first principles package \cite{Giannozzi2009,Giannozzi2017}. The exchange-correlation is described by the BEEF-vdW functional, a bayessian functional designed to treat gas-surface problems and that accounts self-consistently for the van der Waals dispersion~\cite{Wellendorff2012}. The ionic cores are described using ultrasoft pseudopotentials by dal Corso~\cite{dalcorso2014}, specifically accounting for 4, 6, and 10 electrons for C, O, and Pt, respectively.  A plane-wave basis is used to represent the Kohn-Sham states, being the kinetic energy cutoff of 1250 eV. Brillouin zone integration is done by using a $\Gamma$-centered 12$\times$12$\times$1 Monkhorst-Pack mesh~\cite{Monkhorst1976} of special $\mathbf{k}$-points and a Gaussian smearing of 0.27~eV is used for the electronic state occupancies.
All the atoms in the supercell are relaxed until the Hellmann-Feynman forces on each atom are below 2$\times$10$^{-5}~$~Ry/$a_\mathrm{B}$ ($\sim$2$\times$10$^{-4}$~eV/\AA).
The vibrational modes of the system are evaluated using density functional perturbation theory calculations (DFPT). Withing this approach, the dynamical matrices and perturbation potentials are calculated on a $\Gamma$-centered 6$\times$6$\times$1 $\textbf q$ grid.
See Appendixes~\ref{app:Ptslab} and \ref{app:COonPt} for more details about the surface and CO/Pt(111) system, respectively.
We use the electron-phonon Wannier (\textsc{epw}) code to evaluate the  e-ph matrix elements~\cite{NOFFSINGER20102140,PONCE2016116} on the above mentioned $\mathbf{k}$ and $\mathbf{q}$ grids. The \textsc{epw} code further allows to interpolate the matrix elements in finer $\mathbf{k}$ and $\mathbf{q}$ grids taking advantage of the properties of
maximally localized Wannier functions (MLWFs)~\cite{Mostofi2008,Marzari2012,Mostofi2014,Pizzi2020}. 
Within this methodology we represent the electronic structure using a set of 132 MLWFs, constructed using the selected columns of density matrix procedure~\cite{Vitale2020}.

Changes induced in the vibrational spectra due to NC are studied within MBPT by calculating the phonon self-energy due to e-ph coupling $\pi_\lambda(\textbf{q},\omega_{\textbf{q},\lambda})$ ($\lambda$, $\textbf{q}$, and $\omega_{\textbf{q},\lambda}$ are the index, momentum, and energy of the phonon mode, respectively). 
The imaginary part of $\pi_\lambda(\textbf{q},\omega_{\textbf{q},\lambda})$ provides the linewidth [$\gamma_{\textbf{q},\lambda} = -2\mathrm{Im}\pi_\lambda(\textbf{q},\omega_{\textbf{q},\lambda})$], while the renormalization of the phonon frequency is given by the real part. The latter is approximated, provided that the nonadiabatic corrections to the phonon energy are usually small compared to the adiabatic energy by  $\omega_{\textbf{q},\lambda} \approx \omega_\mathrm{A} + \mathrm{Re}\left[\pi_\lambda(\textbf{q},\omega_{\textbf{q},\lambda})-\pi_\lambda(\textbf{q},0)\right]$, with $\omega_\mathrm{A}$ being the adiabatic phonon frequency obtained in the DFPT calculation~\cite{Novko2016,Giustino2017}.

When the CO adlayer is illuminated experimentally with infrared light the excited mode of the CO adlayer is initially vibrating in phase. Thus, we restrict our calculations to the long wavelength limit ($\mathbf{q}=0$) in the phonon self-energy calculations. This is the case of  IRAS experiments and the sum frequency generation (SFG) measurements of Ref.~\cite{Inoue2012}.
Following previous works~\cite{Novko2018,Novko2019,Bombin2023,Bombin2023letter}, we consider that the phonon self-energy consists of first- and second-order terms in the e-ph coupling,  
$\pi_\lambda(\omega_{0,\lambda}) = \pi_\lambda^{[1]}(\omega_{0,\lambda}) + \pi_\lambda^{[2]}(\omega_{0,\lambda})$. The first-order accounts for NC to electron-hole pairs \nolinebreak{(de)excitations}, while the second order corresponds to the so-called EMPPC. 

Similarly  to the Fermi golden rule that has been used by other authors~\cite{hellsing84,Sorbello85, Persson1984, Krishna2006, forsblom07, Askerka2016, maurer16, Diesen21, Schreck22,Hormann2025}, the first-order NC term accounts exclusively for interband electron-hole pair (de)excitations. It reads~\cite{Novko2016, Giustino2017, Novko2018}
\begin{equation}
\pi_\lambda^{[1]}(\omega_{0,\lambda}; T_e)=\sum_{\mu \mu^\prime\textbf{k}\sigma}\left|g^{\mu\mu^\prime}_\lambda(\textbf{k},0)\right|^2\frac{f(\epsilon_{\mu\textbf{k}})-f(\epsilon_{\mu^\prime\textbf{k}})}{\omega_{0,\lambda} + \epsilon_{\mu\textbf{k}}-\epsilon_{\mu^\prime\textbf{k}}+ i\eta},
\label{eq:first_order}
\end{equation}
where $\mu$, $\textbf{k}$, and $\epsilon_{\mu\textbf{k}}$ are the electron band index, momentum, and energy, respectively; $g^{\mu\mu^\prime}_\lambda(\textbf{k},\textbf{q})$ are the e-ph matrix elements; and the summation over $\sigma$ accounts for the spin degree of freedom. We have introduced the Fermi-Dirac distribution function $f(\epsilon_{\mu,\textbf{k}})=1/(e^{\beta[\epsilon_{\mu\textbf{k}}-\mu(T_e)]}+1)$, where $\beta=1/(k_BT_e)$, $k_B$ is the Boltzmann constant, $T_e$ is the electronic temperature, and $\mu(T_e)$ is the chemical potential.
For the calculations presented in this work, we fix $\eta=30$~meV~\cite{Hayashi2013,Schendel2017}, representing the broadening of the electronic states due to scattering processes at the metal~\cite{Novko2016,Rittmeyer17,Bombin2023}. 

The EMPPC term accounts for the coupling between phonons mediated by electrons. This mechanism is important for phonon modes that are decoupled in energy from the other modes of the system, for which the energy mismatch is compensated by the excitation of electron-hole pairs at the metal. 
For low energy phonons, however, the so-called anharmonic coupling~\cite{Persson1984} becomes more important and it collaborates/competes with the e-ph coupling mechanisms, as it has been shown by several authors studying, for example, graphene in different conditions~\cite{Chae2010,Erhardt2025}. In MBPT we account for the EMPPC  by evaluating the second order intraband phonon self-energy that reads~\cite{Novko2018,Novko2019} 
\begin{widetext}
\begin{align}
    \pi_\lambda^{[2]}(\omega_{0,\lambda}; T_e;T_l)=
&-\sum_{\mu \mu^\prime\textbf{k}\sigma,\lambda^\prime\textbf{k}^\prime} \left|g^{\mu\mu}_\lambda(\textbf{k},0)\right|^2
\left[1-\frac{g_\lambda^{\mu^\prime\mu^\prime}(\textbf{k}^\prime,0)}{g_\lambda^{\mu\mu}(\textbf{k},0)}\right]
\left|g^{\mu\mu^\prime}_{\lambda^\prime}(\textbf{k},\textbf{q}^\prime)\right|^2\nonumber\\
&\times\sum_{s,s^\prime=\pm 1}
\frac{f(\epsilon_{\mu,\textbf{k}})-f(\epsilon_{\mu^\prime,\textbf{k}^\prime}-s^\prime s \omega_{\textbf{q}^\prime,\lambda^\prime})}
{\epsilon_{\mu,\textbf{k}}-(\epsilon_{\mu^\prime,\textbf{k}^\prime}-s^\prime s\omega_{\textbf{q}^\prime\lambda^\prime})}
\frac{s\left[n_b(s\omega_{\textbf{q}^\prime\lambda^\prime})+f(s^\prime\epsilon_{\mu^\prime,\textbf{k}^\prime})\right]}{\omega_{0,\lambda}\left[\omega_{0,\lambda} + i\eta +s^\prime(\epsilon_{\mu,\textbf{k}}-\epsilon_{\mu^\prime,\textbf{k}^\prime})+s\omega_{\textbf{q}^\prime\lambda^\prime}\right]},
\label{eq:secon_order}
\end{align}
\end{widetext}
where $\textbf{q}^\prime=\textbf{k}^\prime-\textbf{k}$ and $n_b(\omega_{\textbf{q},\lambda}) = 1/(e^{\beta\omega_{\textbf{q}\lambda}}-1)$ is the Bose-Einstein distribution, with $\beta=1/(k_BT_l)$ and $T_l$ is the lattice (phononic) temperature. 

Equation~\eqref{eq:secon_order} couples the studied ($\textbf{q}\approx$~0, $\lambda$) mode with other ($\textbf{q}^\prime$, $\lambda^\prime$) modes through electron-hole pairs. 
In order to make the calculations computationally feasible, the system is modeled as a six-layer slab. Note, however, that our aim is to reproduce the physics of CO molecules adsorbed at the bulk Pt(111) surface, consequently, we exclude from the phonon self-energy evaluation, the modes that are characteristic of the slab (shear and stretch modes).

In previous works~\cite{Novko2018,Novko2019,Bombin2023letter,Bombin2023}, the vertex correction [second term in the square bracket of Eq.~\eqref{eq:secon_order}] was neglected to reduce the computational cost, as it has been suggested that it introduces a small correction~\cite{Maksimov96,Bauer1998}. Here, we evaluate it by explicitly accounting for the real and imaginary parts of the electron-phonon matrix elements at $\textbf{q}=0$ using a modified version of the EPW package. Note that the matrix elements evaluated at $\mathbf{k}$ and $\mathbf{k}^\prime=-\mathbf{k}$ are related under time-reversal symmetry.

Equations~\eqref{eq:first_order} and \eqref{eq:secon_order} explicitly account for the  electronic $T_e$ and lattice $T_l$ temperatures through the Fermi-Dirac and the Bose-Einstein distributions, respectively. 
This allows us to study the changes in the modes of CO adsorbed on the Pt(111) surface under different conditions. In Sec.~\ref{sec:thermal} we consider thermal conditions, i.e., the $T_e = T_l$ case. Instead, in Sec.~\ref{sec:ultrafast} we consider the ultrafast nonequilibrium conditions (i.e., $T_e \ne T_l$) that are induced when the system is excited with a femtosecond laser pump pulse. To model the latter conditions we rely on a two temperature model~\cite{Anisimov1974}, as it has been described elsewhere~\cite{Germer1993_2,Germer94,Novko2019,Bombin2023}. The model assumes that energy deposited by the laser is initially absorbed by the electronic bath, that is coupled to the phononic bath by the electron-phonon interaction. In Appendix~\ref{app:TTM} we provide further details about the TTM that we employ. As the discussion hereafter focus on the  ($\textbf{q}\approx$~0, $\lambda$) modes, the subscript 0 in the phonon frequency $\omega_{0,\lambda}$ will be omitted, unless otherwise stated.

\section{The system}
\label{subsec:system}
\begin{figure}
    \centering
    \includegraphics[width=1.0\linewidth]{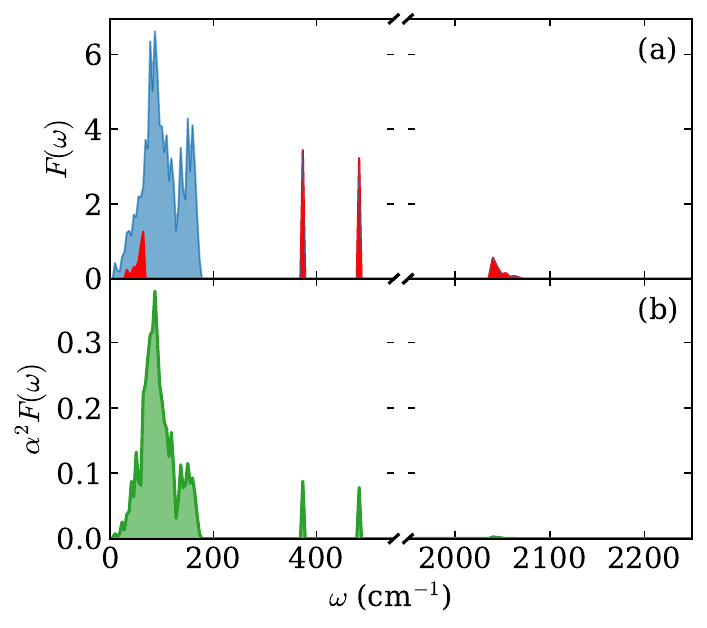}
    \caption{(a) Total phonon density of states (blue) and its projection onto the CO modes (red) (b) Eliashberg function for 0.33~ML CO/Pt(111) at the stable $(\sqrt{3}\times\sqrt{3})$R30$^\circ$-(top) phase. The electron and phonon Brillouin zones are integrated using a $48\times48\times 1$ $\mathbf{k}$ and $\mathbf{q}$ grids with a smearing of 50~meV and 0.10~meV, respectively}
    \label{fig:phonondos}
\end{figure}

Although experimentally characterized since the 1980s, the adsorption structure of CO on Pt(111) has stood as a theoretical challenge for decades due to the failure of DFT in predicting the most stable adsorption position (see discussion in Appendix~\ref{app:COonPt}).
Here we focus on the  $(\sqrt{3}\times\sqrt{3})$R30$^\circ$-(top) adsorption structure that forms at 0.33~ML, which is the coverage studied in the SFG experiments of Refs.~\cite{Inoue2012,Watanabe2010}. Details about this structure are given in Appendix~\ref{app:COonPt}, where we discuss the structural parameters and the corresponding electronic band-structure.

Figure~\ref{fig:phonondos}(a) shows the phonon density of states (ph-DOS) of the 0.33~ML CO/Pt(111) system as well as its projection into the CO adlayer.  
The ph-DOS in the interval $\omega\in[0,190]$~cm$^{-1}$ accounts essentially for the phonon modes of the Pt(111) surface, which in a few cases involve tiny movements of the adsorbed CO. 
The wide peak that is observed at $\omega\in[30,65]$~cm$^{-1}$ in the CO-projected ph-DOS corresponds to the lowest energy modes that can be attributed entirely to the CO adlayer, i.e., the frustrated translation modes (FT). 
As energy increases, three well defined, isolated peaks show up in the spectra. In order of increasing values of $\omega$, they correspond to the frustrated rotation (FR) modes, the CO-Pt(111) mode [hereinafter denoted external stretch (ES)], and the IS mode, respectively. 
The calculated frequencies of the FT, FR, ES, and IS modes at $\textbf{q}=0$ are $\omega_{\rm FT}=34.3$~cm$^{-1}$, $\omega_{\rm FR}=371.5$~cm$^{-1}$, $\omega_{\rm ES}=484.4$~cm$^{-1}$, and $\omega_{\rm IS}=2069.5$~cm$^{-1}$. Note that, at variance with 0.5~ML CO/Pd(111)~\cite{Bombin2023}, the FT, FR, and ES modes in 0.33~ML CO/Pt(111) are energetically well decoupled. We attribute these differences to the different properties of the CO adsorption configuration. In the former, both at hollow and bridge sites, CO is slightly tilted~\cite{Bombin2023}, whereas in the latter, the molecule adsorbs vertically atop a Pt atom (see Appendix~\ref{app:COonPt}).

Figure~\ref{fig:phonondos}(b) depicts the Eliashberg function $\alpha^2F(\omega)$, which measures how strongly coupled is a mode of frequency $\omega$ to the electrons in the system. The results for $F(\omega)$ and $\alpha^2F(\omega)$ show that the shape of the two functions is quite similar for $ \omega<100$~cm$^{-1}$. However, as the phonon frequency increases, the magnitude of the Eliashberg function is substantially reduced compared to $F(\omega)$, pointing to a weaker electron-phonon coupling of these phonons. 
A similar behavior has been observed in other transition metal surfaces; see for example Ref.~\cite{Novko2016} for CO/Cu(100) and Ref.~\cite{Bombin2023} for CO/Pd(111), where the same methodology as here was employed. 

The IS mode of CO adsorbates has been the subject of broad experimental research work in the past because, first, it is infrared active and, second, it is decoupled in energy from the rest of modes in the system.  
These properties have motivated the use of IRAS and other spectroscopy techniques~\cite{Crosslye1977,Krebs1977,Heyden1983,tobin1987,Schweizer1989,Beckerle1991,Carrasco10,Rose2002,Roeterdink2004,kung2000,Watanabe2010,Inoue2012} to elucidate the effect that coverage, adsorption site, temperature, and pressure have on the IS mode properties and to a certain extent in the adsorbate dynamics. 
Interestingly, in the Pt(111) surface, the ES~\cite{baro1979,Hoge1988,Malik1989} and FR~\cite{steininger1982,Lahee1986,Germer1993_2} modes, which are also infrared active and appear isolated in the spectra, have been identified in IRAS experiments.
As a result, the thermal evolution of the ES mode has been characterized by IRAS~\cite{baro1979,Hoge1988,Malik1989}, extending previous studies of the IS mode~\cite{Schweizer1989,Roeterdink2004,Crosslye1977} to a lower energy regime.
In this intermediate-energy region of the vibrational spectra, due to the reduced energy mismatch between the ES mode and the surface and FR modes, three-phonon anharmonic processes are expected to become important~\cite{Ryberg1990,Persson1989}.
Finally, the infrared inactive FT mode 
was measured by inelastic helium atom scattering, being its energy 6~meV (48~cm$^{-1}$)~\cite{Lahee1986}.
Note that the experimentally measured frequencies for the IS, ES, FR, and FT modes at 0.33 ML are in reasonable agreement with our theoretical predictions (see  Table~\ref{table:CO_adsor} in Appendix~\ref{app:COonPt}).

\section{Thermal conditions}
\label{sec:thermal}

In this section we discuss the changes that are induced under thermal conditions ($T$=$T_e$=$T_l$) in the $\textbf{q}$=0 phonon modes of the $(\sqrt{3}\times\sqrt{3})$R30$^\circ$-CO$_\textrm{(top)}$ adlayer formed on Pt(111) at $0.33$~ML coverage. We focus on the IS ($\omega_{\rm IS}=2069.53$~cm$^{-1}$) and ES ($\omega_{\rm ES}=484.34$~cm$^{-1}$) modes, whose thermal changes have been experimentally characterized~\cite{Roeterdink2004,Schweizer1989,Crosslye1977,Malik1989,Hoge1988}. 
Accordingly, we restrict our study to the temperature range $T\in [40,350]$~K, which encompasses the available experimental measurements.  
The first and second order contributions to the frequency and linewidth changes are calculated as $\Delta\omega_{\lambda}^{[1],[2]}(T) \approx \mathrm{Re}[\pi_\lambda^{[1],[2]}(\omega_{\lambda};T)]-\mathrm{Re}[\pi_\lambda^{[1],[2]}(\omega_{\lambda};T=0)$] and $\Delta\gamma_{\lambda}^{[1],[2]}(T) = -2\mathrm{Im}[\pi_\lambda^{[1],[2]}(\omega_{\lambda};T)]+2\mathrm{Im}[\pi_\lambda^{[1],[2]}(\omega_{\lambda};T=0)]$, respectively. In the latter, the superscripts [1] and [2] correspond to the first-order NC and EMPPC mechanisms given by Eqs.~\eqref{eq:first_order} and \eqref{eq:secon_order}, respectively. Note that the total changes $\Delta\omega_{\lambda} (T)$ and $\Delta\gamma_\lambda (T)$ are simply the sum of the two contributions. (For simplicity, the subscript $\lambda$, which specifies the phonon mode, will be omitted, unless it is needed for the discussion).

\begin{figure}[t!]
    \centering
    \includegraphics[width=1.0\linewidth]{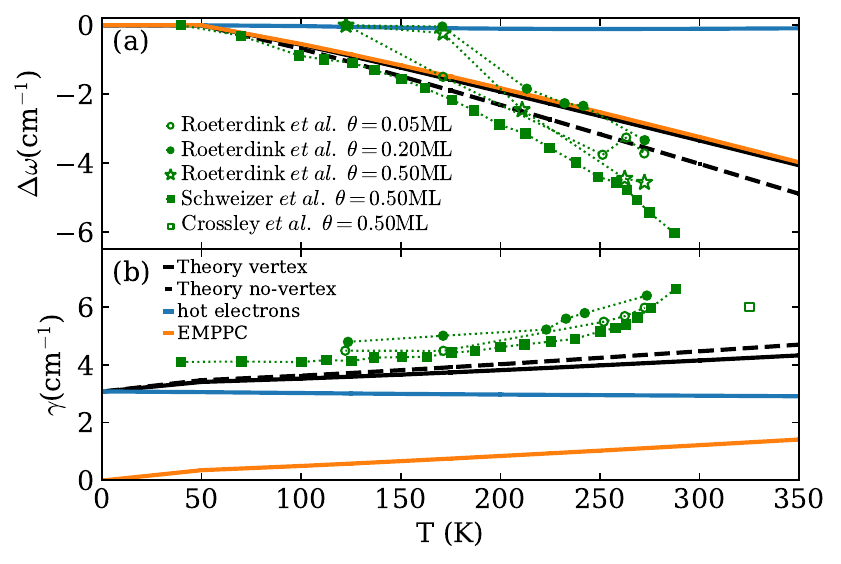}
    \caption{Temperature dependence of the CO IS vibrational mode ($\omega=2069.5$~cm$^{-1}$)
for CO/Pt(111). (a) Frequency shift $\Delta\omega$(T). (b) Total linewidth $\gamma$(T). Black solid curves are the sum of the first-order NC term (blue) plus the vertex-corrected second-order EMPPC term (orange). Black dashed curves are the sum without including the vertex correction. Experimental measurements for different coverages are shown for comparison: $\theta=0.05$~ML (open circle~\cite{Roeterdink2004}), $\theta=0.20$~ML (filled circles~\cite{Roeterdink2004}), and
$\theta=0.50$~ML (filled squares~\cite{Schweizer1989}, open stars~\cite{Roeterdink2004} and open squares~\cite{Crosslye1977}). }
    \label{fig:heatIS}
\end{figure}

Starting with the IS mode, Fig.~\ref{fig:heatIS} shows the corresponding thermal changes arising from electron-phonon coupling up to second order, compared with the experimental IRAS measurements by Crossley \textit{et al.}~\cite{Crosslye1977} and Schweizer \textit{et al.}~\cite{Schweizer1989} and the SFG measurements by Roeterdink \textit{et al.}~\cite{Roeterdink2004}. As shown in panel~(a), our theory predicts a frequency redshift $\Delta\omega=-3.34$~cm$^{-1}$ at $T=300$~K (solid black line). In the same figure, the first-order NC and the second-order (vertex-corrected) EMPPC terms are shown in blue and orange, respectively. For the temperatures considered here, the second-order EMPPC term is responsible for nearly all of the change in frequency, whereas the contribution from the first-order NC term is almost negligible. The results of the frequency shift calculated with (black solid) and without (black dashed) including the vertex correction in EMPPC [see Eq.~\eqref{eq:secon_order}] show that its contribution is small, although it becomes more important as temperature increases. In particular, neglecting the vertex correction results in an overestimation of the frequency redshift by about 1~cm$^{-1}$ at $T=350$~K.

Figure~\ref{fig:heatIS}(b) shows the corresponding results for the thermal changes in the IS linewidth. In this case the effect of the vertex correction within the range of temperatures considered is very minor. In the limit of low temperature ($T<50$~K), our calculations predict a linewidth $\gamma=3.52$ cm$^{-1}$.  The associated lifetime $\tau=1.51$~ps is in the range of the values measured for CO at different metal surfaces, e.g., $1.5\pm0.5$~ps in Pt(100)~\cite{Peremans1995} and $2 \pm 1$~ps in Cu(100)~\cite{Morin1992}. An exception to the previous assertion is the Au(111) surface, where CO is weakly physisorbed and the corresponding IS lifetime is much larger ($49  \pm 3$~ps~\cite{Kumar2019,Novko2019_2}). The IS lifetime of CO adsorbed on Pt(111) was measured by Beckerle \textit{et al.} at various coverages and temperatures~\cite{Beckerle1991}. In particular, at 0.33~ML and $T=150$~K, the authors reported a lifetime of $2.0\pm0.1$~ps ($1.26\pm0.04$~ps) from recovery time (linewidth) measurements that is close to our value at that temperature and coverage ($\tau\approx1.4$ps). From the theoretical side, Maurer \textit{et al.} calculated the CO vibrational lifetimes for 0.25~ML CO/Pt(111) with a $p(2\times2)$-(top) structure~\cite{maurer16}. An IS lifetime $\tau=2.77$~ps was obtained from the first-order NC term calculated within the quasistatic limit [i.e., $\omega_{0,\lambda}\to0$ in Eq.\eqref{eq:first_order}]. 

\begin{figure}
    \centering
    \includegraphics[width=1.0\linewidth]{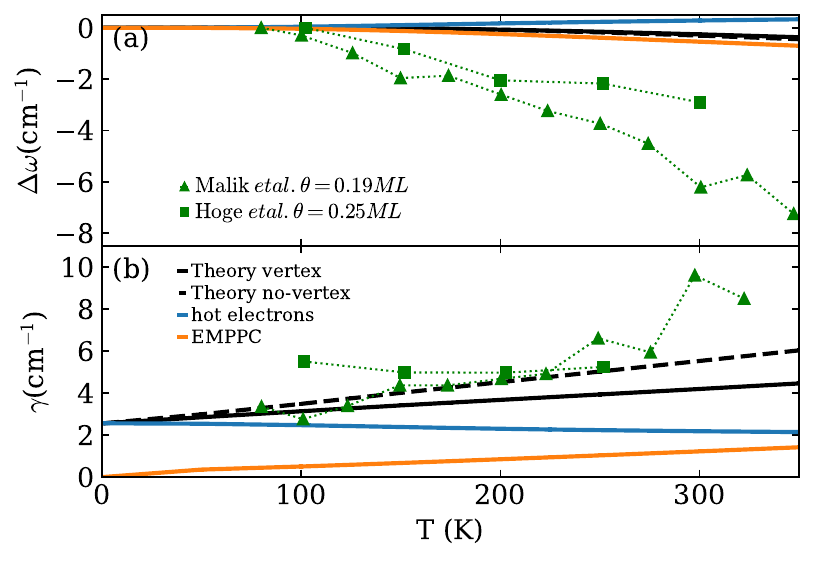}
    \caption{Temperature dependence of the CO ES vibrational mode ($\omega=484.4$~cm$^{-1}$)
for CO/Pt(111). (a) Frequency shift $\Delta\omega$(T). (b) Total linewidth $\gamma$(T). Black solid curves are the sum of the first-order NC term (blue) plus the vertex-corrected second-order EMPPC term (orange). Black dashed curves are the sum without including the vertex correction. Experimental measurements for different coverages are shown for comparison: $\theta=0.19$~ML (triangles~\cite{Malik1989}) and $\theta=0.25$~ML (squares~\cite{Hoge1988}).}
    \label{fig:heatES}
\end{figure}

Figure~\ref{fig:heatIS}(b) shows that the first-order NC contribution (blue curve) dominates the $\gamma(T\to0)$ value. Nonetheless, the change in the linewidth as $T$ increases is dominated by the EMPPC mechanism (orange curve). Note, however, that although the first-order NC contribution to the linewidth thermal change is small, it competes with the EMPPC contribution. At $T=300$~K, for instance, the latter is $\Delta\gamma^{[2]}=1.4$~cm$^{-1}$, whereas the former is $\Delta\gamma^{[1]}\approx-0.2$~cm$^{-1}$, thus, reducing the total linewidth change to 1.2~cm$^{-1}$. 
The fact that the first-order NC term dominates at low temperature and that the thermal changes are driven by the EMPPC has been observed previously for CO/Cu(100)~\cite{Novko2018}  and CO/Pd(111)~\cite{Bombin2023}, using the same formalism as here. In the same panel, we also include the available experimental data for the thermal change of the IS mode linewidth. The experimental measurements at $\theta=0.05, 0.20$, and $0.50$~ML suggest that the effect of coverage on the linewidth change is roughly a constant shift in this temperature range according to the authors~\cite{Roeterdink2004}.  
The experimental broadening is larger than the theoretical one in all the temperature range considered in Fig.~\ref{fig:heatIS}(b). Nonetheless, the theoretical underestimation is somewhat expected because among other things, calculations are performed for a perfect surface with a well defined adlayer structure. 


Let us now focus on the ES mode.
Our theoretical predictions for its thermal changes are shown in Fig.~\ref{fig:heatES}, together with the available experimental data. Similarly to what happens with the IS mode, we observe a redshift in the frequency, accompanied by an increase in the linewidth, as temperature increases. However, the frequency redshift is considerably smaller in the ES mode because, at variance with IS, the EMPPC contribution hardly affects the ES frequency. Note that the value $\Delta\omega=-0.07$~cm$^{-1}$ at $T=300$~K arises from a balance between the small first-order NC and EMPPC contributions, which have opposite signs. 
In contrast, the corresponding thermal increase in the linewidth at $T=300$~K ($\Delta\gamma=1.4$~cm$^{-1}$) is of the order of the one predicted for the IS mode. In the case of $\gamma$, the relative importance of the first-order NC and EMPPC mechanisms is similar to that of the IS mode. First, in the limit of zero temperature the ES linewidth $\gamma=2.54$~cm$^{-1}$ ($\tau=2.10$~ps) is entirely explained in terms of the direct coupling to electron-hole pair (de)excitations. And second, the thermal changes are dominated by the EMPPC mechanism.
Note that although the prediction of $\Delta\gamma$ for the ES mode is close to the measurements by Malik \textit{et al.} at 0.19~ML and in reasonable agreement with those by Hoge \textit{et al.} at 0.25~ML, the experimentally measured $\Delta\omega (T)$ is in both cases much larger than our theoretical prediction. 
Overall, the larger discrepancy between theory and experiment for the ES mode suggests that mechanisms beyond those considered here may play a more important role for this mode than for the IS mode.
As suggested by Persson and coworkers~\cite{Persson1989}, this hints to the increasing importance that the anharmonic coupling between vibrational modes can have in the vibrational relaxation of this intermediate energy mode. Specifically, 
the authors suggested that the coupling to FT is behind the observed changes in the ES frequency shift~\cite{Schweizer1989}. Finally, note that the vertex correction in the ES mode only affects significantly the linewidth change. In particular, it constitutes a reduction of around a 25\% respect to the uncorrected result at the highest temperature.

\section{Ultrafast response}
\label{sec:ultrafast}

\begin{figure*}[]
\centering
    \includegraphics[width=1.0\textwidth]{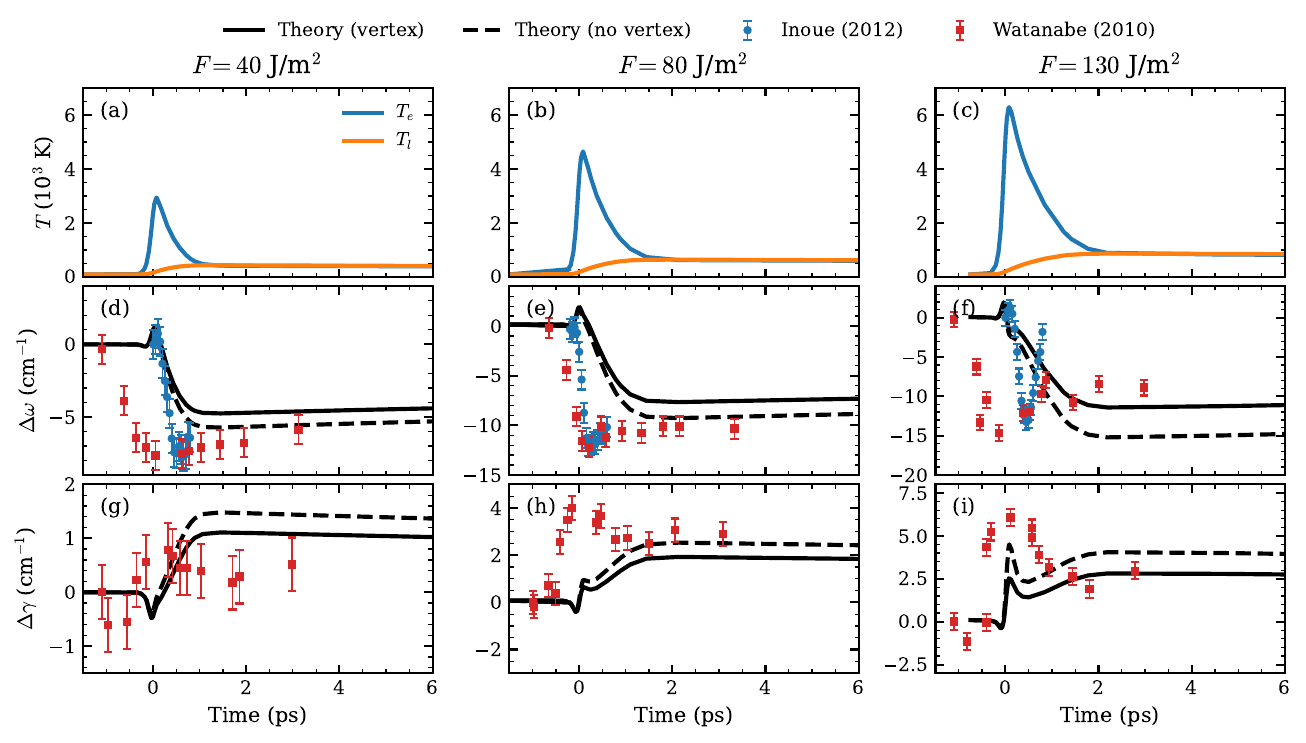}
    \caption{Transient changes induced in the IS vibrational spectra of 0.33~ML CO/Pt(111) by a 400~nm pump pulse with 150~fs duration and absorbed fluence: (left) $F=40$~J/m$^2$, (central) $F=80$~J/m$^2$, and (right) $F=130$~J/m$^2$. The system is initially at $T=90$~K. The peak intensity arrives at $t=0$. (a)-(c) Electron temperature $T_e(t)$ (blue) and lattice temperature $T_l(t)$ (orange) calculated with the TTM. Transient changes in: (d)-(f) the frequency shift and (g)-(i) the linewidth. Black curves correspond to the theoretical prediction without (dashed) and with (solid) vertex correction. Experimental results from Watanabe \textit{et al.}~\cite{Watanabe2010} (red squares) and Inoue \textit{et al.}~\cite{Inoue2012} (blue circles) are included for comparison.}
    \label{fig:theoexpPt}
\end{figure*}

In this section we analyze the ultrafast response of the CO/Pt(111) system excited by an ultrashort laser pulse. More precisely, in order to compare with the experimental results of Refs.~\cite{Watanabe2010,Inoue2012}, we study the time evolution of the frequency shift and linewidth of the IS mode after the system, initially in thermal equilibrium at $T=90$~K, is irradiated by a 400~nm laser pump pulse of 150 fs duration. The excitation generated by the laser pulse is described by the TTM explained in Appendix~\ref{app:TTM}, from which we obtain the time dependent electronic and phononic temperatures that enter Eqs.~\eqref{eq:first_order} and~\eqref{eq:secon_order}. Defining the reference self-energy value $\pi_\lambda(\omega_\lambda;t\to-\infty)$ as the value at $T=90$~K (i.e., before the arrival of the pump-laser pulse), the transient frequency shift and linewidth changes are calculated as $\Delta\omega_\lambda(t) \approx \mathrm{Re}[\pi_\lambda(\omega_\lambda;t)]-\mathrm{Re}[\pi_\lambda(\omega_\lambda;t\to-\infty)]$ and $\Delta\gamma_{\lambda}(t) = -2\mathrm{Im}[\pi_\lambda(\omega_\lambda;t)]+2\mathrm{Im}[\pi_\lambda(\omega_\lambda;t\to-\infty))]$, respectively. The corresponding first- and second-order contributions are obtained as $\Delta\omega_\lambda^{[1],[2]}(t) \approx \mathrm{Re}[\pi_\lambda^{[1],[2]}(\omega_\lambda;t)]-\mathrm{Re}[\pi_\lambda^{[1],[2]}(\omega_\lambda;t\to-\infty))$] and $\Delta\gamma_{\lambda}^{[1],[2]}(t) = -2\mathrm{Im}[\pi_\lambda^{[1],[2]}(\omega_\lambda;t)]+2\mathrm{Im}[\pi_\lambda^{[1],[2]}(\omega_\lambda;t\to-\infty))]$. (In the following, the subscript $\lambda$ will be omitted, unless needed for the discussion.)

In Fig.~\ref{fig:theoexpPt} we compare the results of our calculations with the experimental 
changes reported in Refs.~\cite{Watanabe2010,Inoue2012} from infrared-visible SFG experiments. 
Results for absorbed pump laser fluences $F=40$, 80, and 130~J/m$^2$ are shown in order to compare directly with the experimental data. Figures~\ref{fig:theoexpPt}(a)-~\ref{fig:theoexpPt}(c) show the time evolution of $T_e(t)$ and $T_l(t)$, as obtained from the TTM (see Appendix~\ref{app:TTM} for details). The electronic excitation (described by $T_e$) increases rapidly and reaches its maximum shortly after the arrival of the pump pulse maximum at $t=0$. Next, $T_e$ decreases, but remaining above $T_l$ for around 1~ps at the lowest fluence and 2~ps at the highest one.  
The phononic excitation increases more gradually, reaching its highest $T_l$ value at around 1--2~ps, depending on $F$, and subsequently decays very slowly. This is the usual behavior of the excitation generated by a fs laser pulse in a metal. 

The comparison between theory and experiments for the 
transient frequency shift $\Delta\omega$(t) and linewidth $\Delta\gamma$(t) of the CO IS phonon mode are shown in panels (d)-(f) and (g)-(i) of Fig.~\ref{fig:theoexpPt}, respectively. Regarding the two sets of experimental data~\cite{Watanabe2010,Inoue2012}, both were obtained with the same experimental setup, but using different methods to extract the time evolution of the IS mode characteristics. 
Briefly, in both works, SFG signals were generated by focusing onto the sample a broad--band femtosecond IR pulse and a narrow--band long--duration visible pulse, fixing the time interval between these two pulses at 600 fs. In both cases also, the response of the system to the excitation generated by a 400 nm pump pulse of 150 fs duration was measured. However, in Ref.~\cite{Watanabe2010}, this was achieved by varying the time delay of the IR pulse from the pump pulse and analyzing the measured second-order susceptibility as a function of the pump-IR time delay. This method allowed to obtain both the IS frequency shift and linewidth as a function of time after the arrival of the pump pulse. On the other hand, in Ref.~\cite{Inoue2012} a different strategy was followed. In this case, the pump-IR time delay was fixed to $-0.3$~ps (the IR pulse arrives before the pump pulse). Subsequently, the time domain information was achieved by inverse Fourier transforming the second-order susceptibility measured under these experimental conditions. The analysis of Ref.~\cite{Inoue2012} enabled a higher time resolution. However, only the results for the transient frequency shift but not for the transient linewidth were presented in that reference. 
Additionally, the analyzed time window was limited to 0.8~ps after the pump pulse arrival, whereas in Ref.~\cite{Watanabe2010} the analysis was extended up to 20~ps. Finally, as shown in Fig.~\ref{fig:theoexpPt}, the results reported by the two references for the frequency shift, although similar, are not exactly equal within the common time window, presumably due to the different kind of analysis. Taking all these considerations into account, the degree of agreement between our results and the experimental data from both references can be considered semiquantitative, which is rather satisfactory considering the intricate and nonmonotonic behavior of the transient frequency and linewidth changes.  

Starting with the frequency shift analysis,  our results predict an initial, small but clear blueshift at all fluences, as also found  for CO/Pd(111)~\cite{Bombin2023letter,Bombin2023}. The limited time resolution in Ref.~\cite{Watanabe2010} did not allow to capture this effect. 
In Ref.~\cite{Inoue2012}, although the time resolution was sufficient to resolve the initial blueshift, this feature was not identified, probably, because it was considered an artifact in view of its short duration and its small magnitude. Thus,  the authors reported instead an initial redshift followed by a subsequent blueshift. However, as it can be observed in Fig.~3 of Ref.~\cite{Inoue2012} and, correspondingly, in Fig.~\ref{fig:theoexpPt} here, the initial blueshift is indeed  
present in the experiments at least for the low and high fluences. And, importantly, the duration and magnitude of the peak is perfectly captured by our calculations. This blueshift is  a direct consequence of the particular electronic structure of group-10 transition metals as demonstrated in Refs.~\cite{Bombin2023letter,Bombin2023} for CO/Pd(111) and we discuss below for CO/Pt(111). 
After the initial blueshift, our calculations predict a rapid  redshift, whose magnitude and duration (1--2~ps) increase with the pump fluence. Both facts are in agreement with the experimental observations~\cite{Watanabe2010,Inoue2012}, though in the experiments the redshift takes place faster than in our calculations. Finally, after these 1--2~ps the calculated frequency shift reaches a plateau in agreement with the experimental work that provides results for long times~\cite{Watanabe2010}. In this long time limit, for low and intermediate pump laser fluences, the  
theoretical frequency shift underestimates the experimental value~\cite{Watanabe2010}, but the difference between theoretical and experimental values is rather small in both cases ($\approx 1$~cm$^{-1}$ for $F = 40$~J/m$^2$ and $\approx 2$~cm$^{-1}$ for $F = 80$~J/m$^2$). On the contrary, for the highest fluence ($F=130$~J/m$^2$) our long time results slightly overestimate the experimental frequency shift in $\approx 2$ cm$^{-1}$. 
As a final remark, it should be noted that our results miss 
the rapid blueshift that was observed after the redshift at 130~J/m$^2$ in Ref.~\cite{Inoue2012} and that was ascribed to anharmonic couplings. This effect was neither observed in Ref.~\cite{Watanabe2010}.

Focusing next on the theoretical results for the linewidth changes, Figs.~\ref{fig:theoexpPt}(g)-\ref{fig:theoexpPt}(i) show an initial reduction of the linewidth for the three calculated fluences. Subsequently, the linewidth increases following a somewhat different behavior at low fluence (40~J/m$^2$) compared with intermediate (80~J/m$^2$) and high (130~J/m$^2$) pump laser fluences. At $F=40$~J/m$^2$, the linewidth increases for around 1~ps until it reaches a plateau that extends to long times. At the higher fluences, the transient linewidth change is much more involved. After the already discussed initial decrease, it rapidly increases, peaks at a maximum value at around 0.15~ps, decreases and then rises again to reach the plateau at around 2--3~ps. Both the magnitude of the linewidth change at the 0.15~ps peak and at the long time plateau increase with laser fluence.
Interestingly, this nonmonotonic complex behavior of the linewidth change (its initial decrease, the peak at high fluences and the fluence dependent long time plateau) is also observed in the experiments. Quantitatively, the experimental peaks observed at $F = 80$~J/m$^2$ and $F = 130$~J/m$^2$ are broader and higher than the calculated ones, while the experimental value of the linewidth change at the long time plateau is rather well reproduced by the theory.

The effect of including the vertex correction in the evaluation of Eq.~\eqref{eq:secon_order} is a reduction in the magnitudes of both the frequency and linewidth changes. (Compare the solid and dashed lines in black showing the results with and without the vertex correction.) The correction becomes more important as $F$ increases, which is consistent with its increase with temperature under thermal conditions. In particular, the vertex correction becomes especially pronounced at $F = 130$~J/m$^2$, due to the high values of both $T_l$ and $T_e$ reached under these conditions. For $\Delta\omega$, the long-time value is reduced by about 25\%  respect to the uncorrected value, whereas for $\Delta\gamma$ we obtain a reduction of approximately 30\% at long times (when $T_l$ is high) and up to 45\% at the 0.15~ps peak (i.e., when $T_e$ is high).
Therefore, even if the overall qualitative behavior of the results is not substantially altered, inclusion of the vertex correction can lead to substantial quantitative differences that require further consideration. 

According to Migdal's theorem, vertex corrections are perturbative in the parameter $\lambda\hbar\omega_D/E_{\rm F}$, where $\lambda$ is the electron-phonon coupling constant, $\omega_D$ is the Debye frequency and $E_{\rm F}$ is the Fermi energy~\cite{Migdal1958}. This makes vertex corrections generally negligible for conventional metals. However, the validity of this approximation can be compromised in some systems, including high-temperature superconductors with small Fermi energies~\cite{Pietronero1995,Shuh2002} and materials containing light atoms, such as H and Li, which exhibit high-energy vibrational modes~\cite{Mishra2025,Durajski2016,Szewczyk2020}. Furthermore, systems with flat electronic bands have also been proposed as a setting in which the adiabatic approximation may be broken and where vertex corrections can acquire enhanced importance~\cite{Schrodi2020}. The results for the CO/Pt(111) system studied here point to the limitations of Migdal's approximation also for adsorbates on metal surfaces as well. In this case, the high-energy vibrational modes of the CO adsorbate may contribute to enhancing the role of vertex corrections, in a manner analogous to what has been discussed for materials containing light atoms.

\begin{figure*}[t]
\centering
    \includegraphics[width=1.0\textwidth]{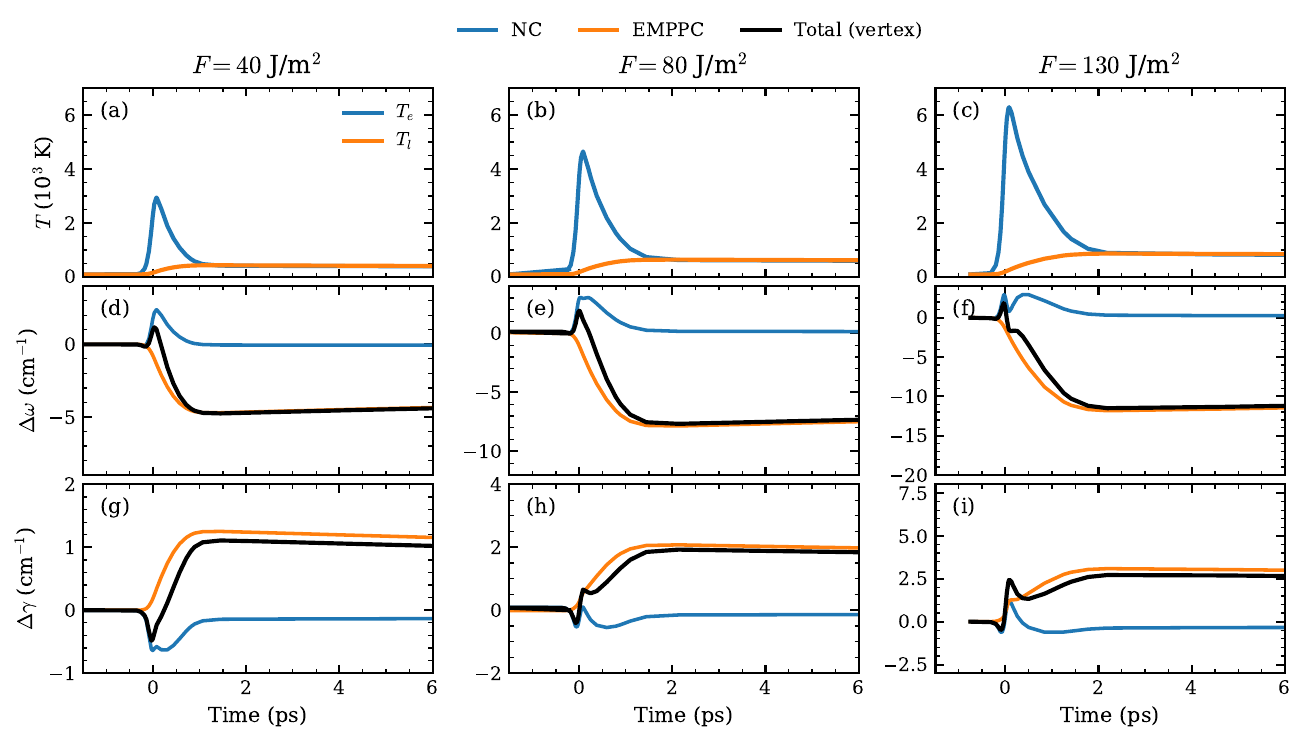}
    \caption{Transient changes induced in the IS vibrational spectra of 0.33~ML CO/Pt(111) by a 400~nm pump pulse with 150~fs duration and absorbed fluence: (left) $F=40$~J/m$^2$, (central) $F=80$~J/m$^2$, and (right) $F=130$~J/m$^2$. The system is initially at $T=90$~K. The peak intensity arrives at $t=0$. (a)-(c) Electron temperature $T_e(t)$ (blue) and lattice temperature $T_l(t)$ (orange) calculated with the TTM. Transient changes in: (d)-(f) the frequency shift and (g)-(i) the linewidth.  Black curves are the sum of the first-order NC term (blue) plus the second-order EMPPC term (orange) including vertex correction. 
    }
    \label{fig:panelTHEO}
\end{figure*}

In order to disentangle the physics behind the above discussed nontrivial time evolution of the frequency shifts and linewidth changes, we show in Fig.~\ref{fig:panelTHEO} the time evolution of the first- and second-order contributions: $\Delta\omega^{[1]}(t)$, $\Delta\gamma^{[1]}(t)$ and $\Delta\omega^{[2]}(t)$, $\Delta\gamma^{[2]}(t)$, as obtained from Eqs.~\eqref{eq:first_order} and \eqref{eq:secon_order}, respectively, for the analyzed pump laser fluences. Regarding the frequency shift, as it was the case in the CO/Pd(111) system~\cite{Bombin2023letter,Bombin2023}, the initial blueshift is created by the first-order NC term or, in other words, is a consequence of the direct coupling to the hot electrons. Note that $\Delta\omega^{[1]}(t)$ is only significant at the very high electronic temperatures (above 1000~K) that are manifested during the first 1--2~ps after the pump laser arrival. The subsequent stronger redshift is due to the second order EMPPC [$\Delta\omega^{[2]}(t)$] and follows the evolution of the lattice temperature. Finally, it is worth to mention the more intricate behavior of $\Delta\omega^{[1]}(t)$ as the pump laser fluence increases. At $F = 40$~J/m$^2$, $\Delta\omega^{[1]}$ just follows the evolution of $T_e$: it increases until $T_e$ achieves its maximum value, where it shows a well defined peak, and then decreases monotonously. At $F = 80$~J/m$^2$, the behavior of $\Delta\omega^{[1]}$ is very similar, though at the maximum of the frequency it shows a plateau that extends for about 250~fs, and even an incipient local minimum seems to exist at the instant of the maximum value of $T_e$. This local minimum at the maximum of $T_e$ is clearly manifested in the case of the highest fluence $F = 130$~J/m$^2$. In this last case, the time evolution of $\Delta\omega^{[1]}$ is rather involved, showing two maxima, the first one when $T_e$ is increasing and a second one, after the local minimum, when $T_e$ decreases. This second maximum is in the region in which $T_l$ has already increased, so that the second order $\Delta\omega^{[2]}$ term almost completely washes it out. All in all, the combination of the two contributions gives rise to the time evolution of the frequency shift described above: a rapid, short and small initial blueshift governed by the coupling of the IS mode with the electrons excited directly by the pump pulse, and a subsequent larger redshift that extends up to long times due to the second order EMPPC process.  

The lower panels of Fig.~\ref{fig:panelTHEO} show how the first-order NC term [$\Delta\gamma^{[1]}(t)$] and the second order EMPPC term [$\Delta\gamma^{[2]}(t)$] contribute to the transient linewidth change. Interestingly, the initial reduction of the linewidth is created by the first-order NC. The first-order NC is also the main responsible of the nontrivial sub-picosecond time evolution of the linewidth at high fluences. In particular, although both first-order NC and EMPPC processes contribute to the increase of the linewidth after its initial reduction, it is due to the time evolution of $\Delta\gamma^{[1]}(t)$ that the peak observed at around 0.15~ps appears. As in the case of the frequency shift, the long time behavior of $\Delta\gamma(t)$ is governed by the EMPPC process that, in this case, constitutes an increase of the linewidth respect to its equilibrium value. 

\begin{figure}
    \centering
    \includegraphics[width=1.0\linewidth]{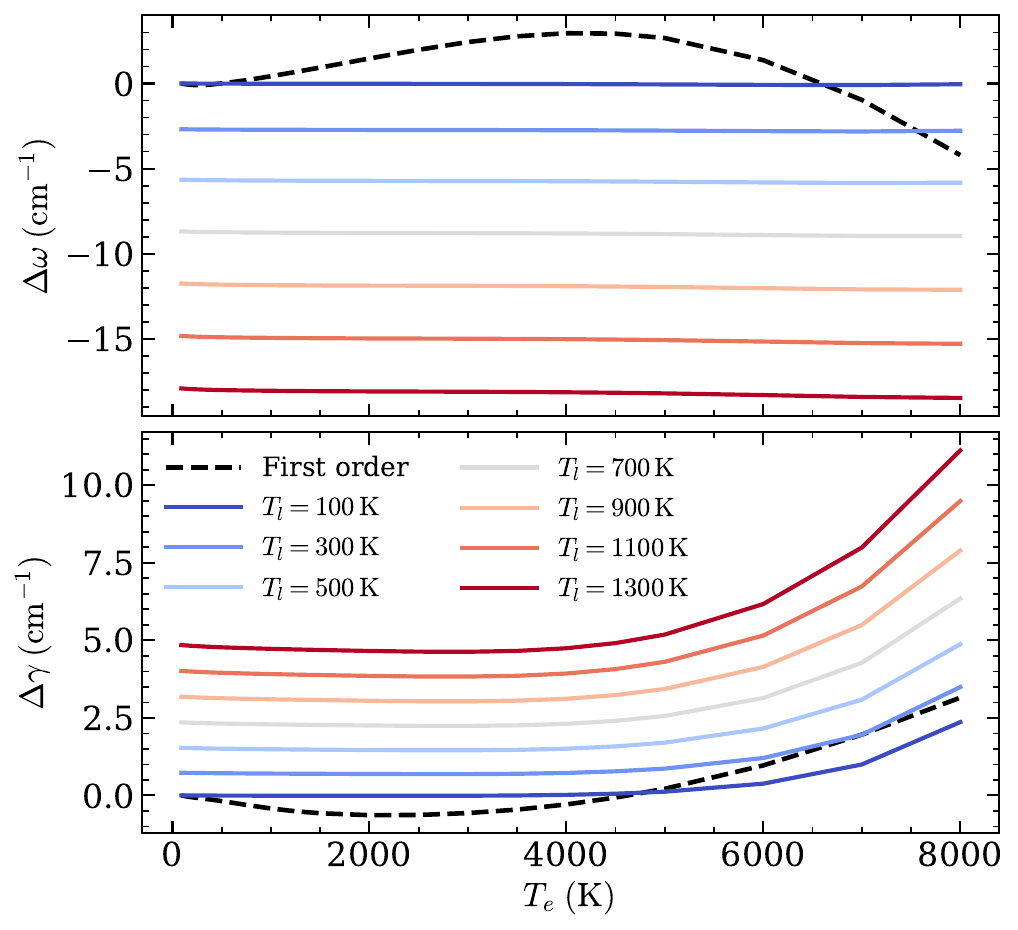}
    \caption{First-order NC (black dashed curve) and EMPPC (colored solid curves) contributions to the frequency (top panel) and linewidth (bottom) change of the IS mode in 0.33~ML CO/Pt(111) as a function of $T_e$. Different EMPPC curves correspond to different $T_l$ as indicated by the legends. Variations are given with respect to the case $T_l = T_e = 90$~K.}
    \label{fig:TeTl}
\end{figure}

In order to understand better the time evolution of the IS mode frequency shift and linewidth change originated by the pump laser pulse, we analyze the dependence of the first-order NC and EMPPC terms on $T_e$ and $T_l$. To this aim, we show in Fig.~\ref{fig:TeTl} the first-order NC and EMPPC contributions to $\Delta\omega$ and $\Delta\gamma$ as a function of $T_e$. In the case of EMPPC, we show results for different values of $T_l$. Note that the temperature ranges considered, $T_e\in[100,8000]$~K and $T_l\in[100,1300]$~K, fully encompass the temperatures achieved under all the studied fluences. 

The top panel of Fig.~\ref{fig:TeTl} shows how the first-order NC contribution to the frequency shift ($\Delta\omega^{[1]}$) produces a blueshift for $T_e>500$~K, whose magnitude increases with $T_e$ up to a maximum value of around 4 cm$^{-1}$ at $T_e=4500$~K. From there on, the magnitude of the blueshift starts to decrease and turns into a redshift for $T_e>6500$~K. The same qualitative dependence of $\Delta\omega^{[1]}$ on $T_e$ was found in the CO/Pd(111) system and the same physical reasons for it apply here~\cite{Bombin2023letter,Bombin2023}. Briefly, the observed behavior is the result of two competing mechanisms. On the one hand, the strength of the e-ph coupling in metals is expected to be enhanced as $T_e$ increases, which would lead to a redshift due to the first-order NC term. This is what is obtained, for instance, in the case of a coinage metal surface as Cu(100)~\cite{Novko2019,Inoue2016}. On the other hand, as it was demonstrated for the case of Pd(111)~\cite{Bombin2023letter}, due to the abrupt reduction of the electronic density of states above the Fermi energy, the number of electrons contributing to the e-ph interaction diminishes with increasing $T_e$. This effect effectively weakens the e-ph interaction leading to the observed blueshift. This latter mechanism initially dominates as $T_e$ increases; whereas only at extremely high electronic temperatures-- $T_e>6500$~K in the case of CO/Pt(111)--, the former mechanism becomes dominant and a redshift is observed. 

The $T_e$ dependence of $\Delta\omega^{[1]}$ fully explains its time evolution for the different pump laser fluences that was discussed in Fig.~\ref{fig:panelTHEO}. For the low fluence $F = 40$~J/m$^2$, the highest $T_e$ remains clearly below 4500~K and, therefore, it does not reach the temperature at which $\Delta\omega^{[1]}$ starts to decrease. As a result, for this fluence $\Delta\omega^{[1]}(t)$ increases until the highest $T_e$ is reached and then decays. For $F=80$~J/m$^2$, $T_e(t)$ barely surpasses the value of 4500~K 
and, as a result, we observe the above discussed plateau during the time period at which $T_e$ remains above 4000~K. Lastly, the intricate behavior of $\Delta\omega^{[1]}(t)$ at $F=130$~J/m$^2$ in Fig.~\ref{fig:panelTHEO}(f) is explained by the fact that in this case the highest $T_e$ is higher than 6000~K, i.e., well beyond the $T_e$ at which $\Delta\omega^{[1]}(T_e)$ in Fig.~\ref{fig:TeTl} is maximum. This explains that for this fluence $\Delta\omega^{[1]}(t)$ starts increasing until the time instant at which $T_e \approx 4500$~K (first peak),  
then decreases with time until reaching the $T_e(t)$ maximum (local minimum), it again increases as $T_e(t)$ again decreases down to $\approx$~4500~K (second peak), and it finally decreases with decreasing $T_e(t)$. Regarding the EMPPC mechanism, the upper panel of Fig.~\ref{fig:TeTl} shows that $\Delta\omega^{[2]}$ is roughly independent of $T_e$ and that induces a redshift whose magnitude increases with $T_l$. This explains why the magnitude of the redshift increases with the value of the pump laser fluence, and also why the time evolution of $\Delta\omega^{[2]}$ follows the time evolution of $T_l$ as shown in Fig.~\ref{fig:panelTHEO}. 

The bottom panel of Fig.~\ref{fig:TeTl} shows the $T_e$ dependence of the first-order NC and EMPPC contributions to the linewidth change. In the case of EMPPC, this dependence is calculated at different $T_l$. Initially, as $T_e$ increases the first-order NC term $\Delta\gamma^{[1]}$ becomes negative, i.e., $\gamma^{[1]}$ decreases respect to its value at $T=90$~K. It remains negative up to 
$T_e \approx$ 4500~K, reaching a maximum reduction of $\approx 0.6$ cm$^{-1}$ at about 2000~K. For $T_e >$~4500~K, $\Delta\gamma^{[1]}$ becomes positive and its value increases with $T_e$. This behavior explains the time evolution of $\Delta\gamma^{[1]}(t)$ that was shown in Fig.~\ref{fig:panelTHEO} for the different laser fluences. At $F=40$~J/m$^2$, $T_e$ is always smaller than 4500~K and, therefore, the first-order NC term induces a reduction in the linewidth at all times. 
The nearly constant value of $\Delta\gamma^{[1]}(t)$ during the approximately first 0.5~ps, over which we observe the maximum linewidth reduction, can also be understood by considering that, during this time period, $T_e$ remains in the range of 1500--3000~K, for which the variation of $\Delta\gamma^{[1]}$ is very small. The more complicated behavior of $\Delta\gamma^{[1]}$ at higher fluences ($F=80$~J/m$^2$ and $F=130$~J/m$^2$) is due the higher $T_e$ induced in the system, which include the $T_e$ range for which, after its initial reduction, $\Delta\gamma^{[1]}(T_e)$ increases. In particular, the above discussed 0.15~ps peak observed in $\gamma^{[1]} (t)$ for the highest fluence $F=130$~J/m$^2$, occurs because $T_e$ achieves values for which $\Delta\gamma^{[1]}$ becomes positive.

The contribution of the EMPPC mechanism $\Delta\gamma^{[2]}$ increases with $T_l$ and is, thus, responsible of the long time behavior ($t>2$~ps) and its fluence dependence. As Fig.~\ref{fig:TeTl} shows, $\Delta\gamma^{[2]}$ remains approximately independent of $T_e$ up to a threshold of about 5000--6000~K that depends on $T_l$. More precisely, the electronic temperature at which a noticeable enhancement of $\Delta\gamma^{[2]}$ starts to take place, decreases as $T_l$ increases, from approximately 6000~K at $T_l=$~100~K to 5000~K at $T_l=$~1300~K. As a consequence, for the range of $T_e$ and $T_l$ values that are induced in the system at $F=40$ and 80~J/m$^2$, the $T_e$ dependence of $\Delta\gamma^{[2]}$ does not play any noticeable role in the transient linewidth change. However, for $F=130$~J/m$^2$, the initial steep increase in the EMPPC contribution to the transient linewidth up to the 0.15~ps peak [see $\Delta\gamma^{[2]}(t)$ in Fig.~\ref{fig:panelTHEO}(i)] is partly attributable to the $T_e$ dependence of EMPPC.

Finally, from a broader perspective, our results on the transient linewidth change highlight the crucial role of the electronic temperature in modulating the electron-phonon coupling strength, which ultimately dictates the energy exchange and, hence, the dynamics of adsorbates on metal surfaces~\cite{Bombin2026}.

\section{Conclusions}
\label{sec:conclusions} 

We have studied the vibrational dynamics of CO adsorbed on Pt(111) at a coverage of 0.33~ML using a first-principles framework that combines many-body perturbation theory with density functional perturbation theory. This work is motivated by two main considerations. First, the CO/Pt(111) system has been extensively investigated experimentally, both under thermal equilibrium and under ultrafast nonequilibrium conditions, providing an ideal benchmark for our theoretical approach. Second, some of us recently applied the same methodology as here to the CO/Pd(111) system and predicted the emergence of a transient blueshift of the internal stretch mode after excitation with a femtosecond laser pulse~\cite{Bombin2023letter,Bombin2023}. Although we showed that this effect originates primarily from the electronic structure of the Pd(111) surface, the lack of suitable experimental data prevented a direct validation of the proposed mechanism. Since Pd and Pt are both group 10 transition metals with similar electronic structures, the present study provides us the opportunity to benchmark the proposed mechanism for the origin of the blueshift~\cite{Bombin2023,Bombin2023letter} against the available pump-probe measurements for CO/Pt(111).

We have first applied our methodology to study the thermal changes induced in the internal (IS) and external stretch (ES) modes under thermal heating. For both modes a redshift is induced, accompanied by an increase in the linewidth. 
In each mode, the magnitude of the linewidth at low temperature and its thermal evolution are dominated by the nonadiabatic electron-hole pair excitations (first-order NC) and by the electron-mediated phonon-phonon excitations (EMPPC), respectively, as previously obtained for CO adsorbed on Cu(100)~\cite{Novko2019} and Pd(111)~\cite{Bombin2023letter}.
These findings  are in reasonable agreement with the experimental measurements.
However, when analyzing the thermal frequency shift $\Delta\omega(T)$, differences between the two vibrational modes emerge. 
For the IS mode, the $\Delta\omega(T)$ prediction is in reasonable agreement with the low-coverage measurements for CO adsorbed at top sites. However, for the ES mode, the predicted frequency shift is much smaller than in experiments. 
The different accuracy of the predictions for the IS and ES modes highlights, on the one hand, the importance of EMPPC in the renormalization of high-energy modes, such as the IS mode, and, on the other hand, the increasing importance of anharmonic phonon-phonon coupling for lower-energy modes~\cite{Persson1989}.

When the 0.33~ML CO/Pt(111) system is irradiated with a femtosecond laser pulse, the IS mode exhibits transient changes in the time scale of femtoseconds (picoseconds) that  are dominated by the coupling to electrons (to other phonons mediated by electrons). As previously found for CO/Pd(111), the main finding is the emergence of a transient blueshift in the femtosecond regime, driven by the characteristic crossing of the Fermi level by the d-band edge. This initial blueshift is accompanied in CO/Pt(111) by a transient reduction of the linewidth. Remarkably, both features were also present in the experimental data, although they were not explicitly discussed.
The magnitude of the transient changes that are measured in the IS frequency and linewidth after 2--4~ps of the pump-pulse excitation is in fairly good agreement with our predictions. 

In this work, we have introduced a methodological improvement, not included in previous studies~\cite{Novko2018,Novko2019,Bombin2023,Bombin2023letter}, that consists in the explicit evaluation of the vertex-correction screening in the electron-mediated phonon-phonon coupling mechanism. In the CO/Pt(111) system, including the vertex correction decreases the magnitude of both the frequency and linewidth changes with temperature. The effect is particularly  important in the nonequilibrium conditions generated under laser irradiation due to the high temperatures reached. In this case, we obtain large reductions of both the frequency shift and linewidth change of the IS mode that, at high laser fluences, can amount up to about 25\% and 45\%, respectively. These results point to potential limitations of Migdal's approximation in describing high-energy vibrational modes of adsorbates on metal surfaces, in line with existing studies demonstrating similar limitations in other contexts that involved materials containing light atoms~\cite{Mishra2025,Durajski2016,Szewczyk2020}.

\acknowledgments{
Financial support was provided by the
Spanish MCIN$/$AEI$/$10.13039$/$501100011033$/$, FEDER Una manera de hacer Europa (Grant No.~PID2022-140163NB-I00), Gobierno Vasco-UPV/EHU (Project No.~IT2032-26), and the Basque Government Education Departments’ IKUR program, also co-funded by the European NextGenerationEU action through the Spanish Plan de Recuperación, Transformación y Resiliencia (PRTR). Computer resources were provided by the Donostia International Physics Center (DIPC) Supercomputing Center. 
}

\appendix

\begin{table*}[tb!]
\caption{CO properties of the optimized  $\sqrt{3}\times\sqrt{3}$R30$^\circ$-(top) equilibrium structure corresponding to 0.33~ML of CO adsorbed atop on Pt(111):  Adsorption energy per molecule $E_a$, bond length $d$(C--O), adsorption height of the C atom $Z$ (respect to the Pt topmost layer), and IS, ES, FR, and FT ($\textbf{q}=0$) phonon frequencies $\omega_{0}$.}
\label{table:CO_adsor}
\begin{ruledtabular}
\begin{tabular}{ l c c c c c c c }
& $E_a$ (eV) & $d$(C--O) (\AA) & $Z$ (\AA) &
$\omega_{0}^{\rm IS}$ (cm$^{-1}$) &
$\omega_{0}^{\rm ES}$ (cm$^{-1}$) &
$\omega_{0}^{\rm FR}$ (cm$^{-1}$) &
$\omega_{0}^{\rm FT}$ (cm$^{-1}$) \\
This work & -1.37 & 1.15 & 2.06 & 2069.5 & 484.4 & 371.5 & 34.3\\
Experiments & -1.17~\cite{Ertl1977} & 1.15$\pm$0.10~\cite{Blackman1988} & 1.84$\pm$0.10~\cite{Blackman1988} & 2100~\cite{Schweizer1989,Crosslye1977}  &475~\cite{Malik1989}    & 411~\cite{steininger1982,Lahee1986}   & 48.5~\cite{Lahee1986}  \\
            & -1.55~\cite{Yeo1997}  &               &                 & 2090~\cite{Roeterdink2004,kundappaden2019dft}       & 480~\cite{baro1979} & 350~\cite{baro1979}\\
            &                       &               &                 & 2104~\cite{Inoue2012,Watanabe2010}                      & 474~\cite{Hoge1988}    \\
            &                       
\end{tabular}
\end{ruledtabular}
\end{table*}

\section{Platinum surface}
\label{app:Ptslab}
The Pt(111) surface is modeled with a $\sqrt{3}\times\sqrt{3}$R30 surface unit cell that includes six Pt-layers with 14.0~{\AA} of vacuum.  The latter is the minimum cell that allows us to study  the experimental CO coverage of 0.33~ML (see Appendix~\ref{app:COonPt} for details about the CO/Pt(111) structure). The relaxed structure is obtained by minimizing the atomic force components below 10$^{-5}$~Ry/$a_\mathrm{B}$. After relaxation the theoretical lattice constant is $a=4.0$~{\AA}, which compares well with the experimental value of  $3.92$~\AA~\cite{Ashcroft1988}. The theoretical interlayer distances between the $ith$ and $jth$ layers $d_{i,j}$ are: $d_{1,2}=d_{5,6}=2.337$~{\AA}, $d_{2,3}=d_{4,5}=2.305$~{\AA}, and $d_{3,4}=2.308$~{\AA}. Note that the central interlayer distance is almost equal to the one expected in bulk Pt ($a/\sqrt{3}=2.309$). Existing measurements of the first and second interlayer distances in a Pt(111) surface reported values: $d_{1,2}=2.29\pm0.01$~{\AA}~\cite{Materer1995} and $2.285\pm0.01$~{\AA}~\cite{Gallego1997} and $d_{2,3}=2.27\pm0.03$~{\AA}~\cite{Materer1995} and $2.245\pm0.013$~{\AA}~\cite{Gallego1997}. Thus, the predictions of the BEEF-vdW functional for the expansion and contraction of the first and second interlayer distances agree with the experimental measurements within the experimental error.

\section{CO adsorption structure}
\label{app:COonPt}

Experimentally, several adsorption structures have been identified that depend on the CO coverage. 
In the low coverage limit, CO adsorbs upright at top positions. As the coverage $\theta$ is increased in the range $\theta<0.33$~ML, CO molecules form islands characterized by a dominant $(\sqrt{3}\times\sqrt{3})$R30$^\circ$-(top) phase in the central parts of these islands and the presence of CO adsorbed on bridge sites in the borders. 
At 0.33~ML, the $(\sqrt{3}\times\sqrt{3})$R30$^\circ$-(top) phase is formed.
For $0.33 < \theta < 0.50$~ML, bridge occupation increases until the uniform $c(4\times2)$-(bridge-top) phase is reached. 
The latter was well established in the 70s and 80s by electron energy-loss 
spectroscopy~\cite{Froitzheim1977,hopster1978,baro1979,steininger1982,Blackman1988}, low energy electron 
diffraction~\cite{Ertl1977,Ogletree1986,steininger1982}, IRAS~\cite{Krebs1977,Crosslye1977,horn1977,Heyden1983,Malik1989,Hoge1988,Ryberg1990,Schweizer1989,tobin1987}, X-ray photoelectron spectroscopy~\cite{Norton1979}, angle resolved photoemision spectroscopy~\cite{trenary1983} and inelastic helium scattering~\cite{Lahee1986}.
More recently, this picture has been confirmed by in situ scanning tunneling microscope images~\cite{Yang2013}.

However, the adsorption structure of CO/Pt(111) has long posed a theoretical challenge. 
In the late 90s, it became manifest the failure of LDA/GGA DFT in predicting the correct adsorption structure of CO on Pt, which has been extensively referred in the literature as the CO \textit{puzzle} after the work by Feibelman \textit{et al.}~\cite{feibelman2001co}.
Conventional (LDA/GGA) DFT calculations predict the hollow site as the most stable adsorption site. In the early 2000s, numerous studies were trying to understand this adsorption site problem (see Refs.~\cite{feibelman2001co, olsen2003co, kresse2003significance} to cite some). Using the vdw-DF exchange-correlation functional by Dion \textit{et al.} that includes vdW dispersion in a self-consistent manner~\cite{dion2004van}, Lazi\'{c} \textit{et al.} showed that inclusion of the nonlocal correlation solved the CO puzzle on different surfaces, but not on Pt(111)~\cite{lazic10}.  Later on,  in a comparative study including 30 (GGA and meta-GGA) exchange-correlation functionals, Janthon \textit{et al.}  showed that
only the M06-L functional~\cite{zhao2006new} provides the correct relative stability between the top and hollow sites on Pt(111)~\cite{janthon2017}. However, M06-L is known to inadequately account for dispersion interactions~\cite{Goerigk2011} and the authors also showed in the same study that the dispersion-corrected M06-L predicted instead the hollow site as the stable adsorption site, that hints to a cancellation of errors in the M06-L energies. 
In 2018, Gunasooriya \textit{et al.} used vdW-DF to study the thermodynamic adsorption stability of CO on Pt(111) for coverages ranging from isolated CO to ordered high-covered structures~\cite{gunasooriya2018co}.  Their calculations were able to correctly predict the increase and decrease in the bridge:top ratio with coverage, from 0 to 1:1 and back to 0. 
In a more recent study, Kundappaden \textit{et al.}  \cite{kundappaden2019dft} compared different van der Waals
functionals and showed that only vdW-DF and vdW-DF2~\cite{murray2009} provide the correct preference for the adsorption site, but with a small energy difference. It has been shown that the CO puzzle is also solved when more accurate many-body techniques are employed to evaluate the correlation energy such as the random phase approximation~\cite{Schimka2010} and quantum Monte Carlo~\cite{Hsing2019}.

In this work, following our previous study for the CO/Pd(111) system,  we have chosen to  use the BEEF-vdW functional that includes vdW dispersion in vdW-DF2 fashion. 
Using this functional and the computational setup explained in Sec.~\ref{sec:methods}, the adsorption energy per molecule for the $\sqrt{3}\times\sqrt{3}$R30$^\circ$ structure with CO adsorbed atop is $-1.37$~eV, which is 10~meV more stable than adsorption on bridge. The atop value compares well with the measured adsorption energy per molecule of 1.17eV~\cite{Ertl1977} and 1.55~eV~\cite{Yeo1997} that were reported for the same structure. 
In Table~\ref{table:CO_adsor} we summarize the properties of the CO molecule adsorbed on Pt(111) with the $\sqrt{3}\times\sqrt{3}$R30$^\circ$-(top) equilibrium structure.

\begin{figure}[tb]
    \centering
    \includegraphics[width=1.0\linewidth]{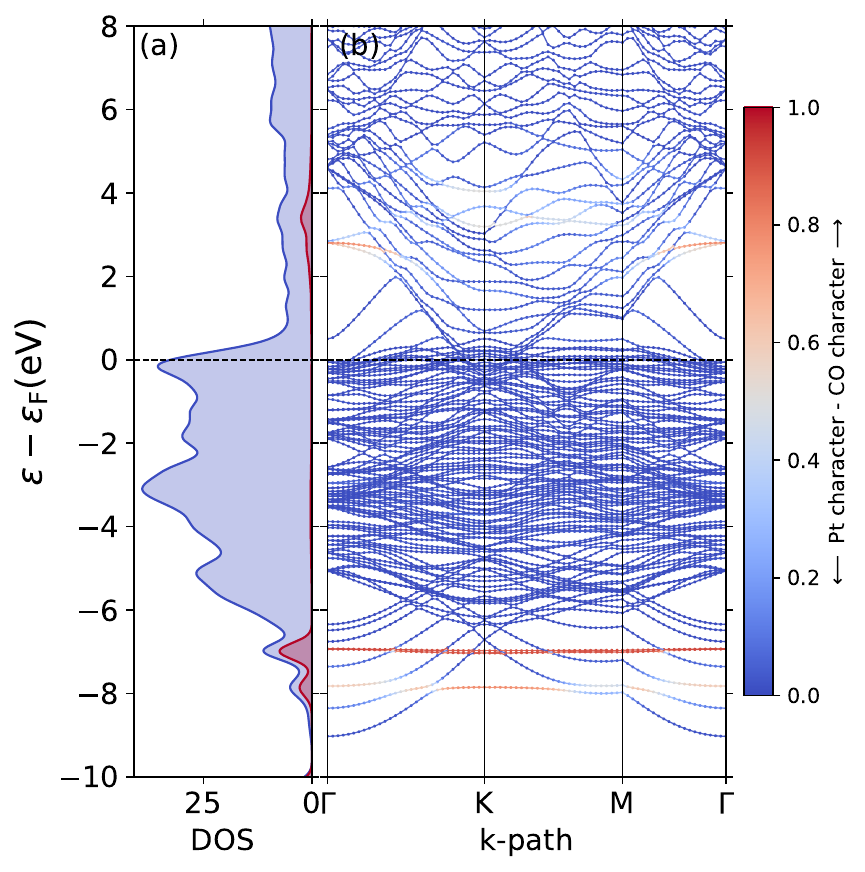}
    \caption{(a) Electronic density of states for 0.33~ML CO/Pt(111) (blue), showing also the projection onto C and O atomic orbitals (red).
    (b) Corresponding band-structure. The color bar encodes the CO (reddish) or Pt (bluish) character for each band at a given $\mathbf{k}$-point. 
    Dashed black lines denote the Fermi level.}
    \label{fig:bandstructureCOPt}
\end{figure}

The electronic density of states (DOS) and corresponding bandstructure are shown in Fig.~\ref{fig:bandstructureCOPt}. The chosen energy range accounts for all the bands that are expected to have electron populations different from 1 or 0, at the extreme temperatures that are considered in Sec.~\ref{sec:ultrafast}, where we study the nonequilibrium dynamics after pumping the system with an ultrashort laser pulse. The DOS is considerably larger at $\varepsilon <E_F$ than above $E_F$ due to the the presence of the $d$-band edge around $E_F$. Thus, as temperature increases, the chemical potential of the system shifts to larger values, ensuring conservation of the total number of electrons in the system. In this respect, the electronic structure of 0.33~ML CO/Pt(111) and 0.50~ML CO/Pd(111)~\cite{Bombin2023} are qualitatively similar, explaining the predicted initial blueshift of the IS CO mode frequency in pump-probe experiments. However, some differences can also be identified. The $5d$ bands of Pt span over a larger energy range than the $4d$ bands of Pd, resulting in a somewhat less sharp $d$-band edge in Pt. The occupied $\sigma$ and $\pi$ bonding orbitals of CO hybridize more strongly with the $sp$-bands of Pt than with those of CO/Pd(111)~\cite{Bombin2023}, whereas a similar hybridization of the CO antibonding orbitals that lie about 3~eV above $E_F$ is obtained in both systems.

\section{Two-temperature model}
\label{app:TTM}

The two-temperature model (TTM)~\cite{Anisimov1974} provides a description of the energy balance between a bath of phonons interacting with a bath of electrons that are excited by a laser pulse. As detailed elsewhere~\cite{frischkorncr06, saalfrank2006}, mathematically it reads, 
\begin{equation}
\begin{split}\label{eq:TTM}
    C_\mathrm{e}\frac{\partial T_\mathrm{e}}{\partial t} &=
        \frac{\partial}{\partial z}\left(\kappa_\mathrm{e}\frac{\partial T_\mathrm{e}}{\partial z}\right)-G(T_\mathrm{e}-T_\mathrm{l})+S\\
    C_\mathrm{l}\frac{\partial T_\mathrm{l}}{\partial t} &=
        G(T_\mathrm{e}-T_\mathrm{l}),
\end{split}
\end{equation}
where $C_\mathrm{e}(T_e)$ and $C_\mathrm{l}(T_l)$ are the electron and lattice heat capacities, respectively; $\kappa_\mathrm{e}$ is the thermal conductivity of the electron subsystem used in its simplified form~\cite{Ashcroft1988} (which is a reasonable approximation for $d$-metals~\cite{Petrov2013}): $\kappa_\mathrm{e}(T_\mathrm{e},T_\mathrm{l})=\kappa_0\frac{T_\mathrm{e}}{T_\mathrm{l}}$,  with $\kappa_0=  76.2$ W/(m K)~\cite{Germer1993_2} being the room temperature (300~K) thermal conductivity of platinum; $z$ denotes the spatial direction perpendicular to the surface plane; and $S$ is the laser source term.
The $T_e$-dependent values of $C_e(T_e)$ and $G(T_e)$ for Pt are taken from the \textit{ab-initio} calculations by Lin \textit{et al.}~\cite{Lin2008}. The value of $C_l(T_l)$ is taken from experimental measurements in the range $T_l \in [80, 1000]$~K, and extrapolated linearly to higher temperatures~\cite{Yokokawa1979}.

In our TTM, the energy of the (pump) laser pulse is absorbed by the electronic bath at $z=0$ and $t=0$, and its spatial and temporal profile is modeled with the following form
\begin{equation}
S(z,t) = \frac{a_{\lambda}F}{2\Delta t}e^{-\alpha_\lambda z} {\rm sech}^2\left(\frac{t}{\Delta t}\right)
   \label{eqn:TTMpulse} 
\end{equation}
where $\alpha_\lambda=4\pi\kappa_\lambda/\lambda$ is the light absorption coefficient of the metal for a source of wavelength $\lambda$, $\Delta t$ the full width at half maximum of the temporal profile of the laser pulse, and $F$ the absorbed fluence.
The parameters of the pulse are taken to reproduce the 400~nm, 150~fs laser pulse of Ref.~\cite{Watanabe2010} and $\kappa_{{400{\rm nm}}}=2.8472$~\cite{Rakic1998} is the light extinction coefficient of Pt at the experimental wavelength.

\bibliography{refs}

\end{document}